\documentclass[12pt]{article}
\usepackage{amsmath,amssymb,amsfonts,amsthm}

\title{A new class of $f$-deformed charge coherent states and their nonclassical properties}

\author{ M Mortazavi$^{1}$, M K Tavassoly$^{1, 2}$
\\
\footnotesize{1- Department of Mathematical Sciences, Yazd University, Yazd, Iran} \\ \footnotesize{2- Research Group of Optics and Photonics,
Yazd University, Yazd, Iran}
\\ \footnotesize{E-mail: mktavassoly@yazduni.ac.ir;  }}

\begin{document}

\maketitle

  \newcommand{\HD}{\mathds{H}}
 \begin{abstract}
Two-mode charge (pair) coherent states has been introduced
previously by using $\langle\eta|$ representation. In the present
paper we reobtain these states by a rather different method. Then,
using the nonlinear coherent states approach and based on a
simple manner by which the representation of two-mode
charge coherent states is introduced, we generalize the bosonic creation and
annihilation operators to the $f$-deformed ladder operators and
construct a new class of $f$-deformed charge coherent states.
Unlike the (linear) pair coherent states, our presented structure has the potentiality to generate
a large class of pair coherent states with various nonclassicality signs and physical properties which are of interest.
Along this purpose, we use a few well-known nonlinearity functions associated with particular
quantum systems as some physical appearances of our presented
formalism.  After introducing the explicit form of the above
correlated states in two-mode Fock-space, several
nonclassicality features of the corresponding states (as well as the two-mode
linear charge coherent states) are numerically investigated by calculating quadrature squeezing, Mandel
parameter, second-order correlation function, second-order
correlation function between the two modes and Cauchy-Schwartz
inequality. Also, the oscillatory behaviour of the photon
count and the quasi-probability (Husimi) function of
the associated states will be discussed.
\end{abstract}

 {\bf Pacs: 42.50.Dv, 42.50.-p}

 \vspace{2pc}

 \section{Introduction}\label{sec-intro}
 Coherent states are venerable objects in many areas of physical researches in recent decades \cite{Klauder, Ali}, with special place in the quantum optics \cite{Gazeau, perelomov}.
 Therefore, various generalizations have been proposed up to now.
 In this paper, we confine ourselves to two-mode type of generalization of coherent states. Along this latter subject,
 we start with a brief review on the charge (pair) coherent states.
 Horn and Silver \cite{Horn} defined the so-called charge coherent state $ |\alpha,q\rangle$, which is the common eigenvector
 of number difference operator, sometimes has been named charge operator, given by
 \begin{eqnarray}\label{Q,BAR}
\hat{Q}= \hat{a^{\dag}}  \hat{a}- \hat{b^{\dag}} \hat{b}
\end{eqnarray}
and pair annihilation operator $\hat{a}\hat{b}$ with eigenvalues $q$ and $\alpha$, respectively, i.e.,
\begin{eqnarray}\label{aQ1}
 \hat{Q}|\alpha,q\rangle=q|\alpha,q\rangle,
 \end{eqnarray}
 \begin{eqnarray}\label{aab1}
 \hat{a}\hat{b}|\alpha,q\rangle=\alpha|\alpha,q\rangle,
 \end{eqnarray}
 where $\hat{a}(\hat{a^{\dag}})$ and $\hat{b}(\hat{b^{\dag}})$ are bosonic annihilation (creation) operators and $q$ is an integer has been named $"$charge number$"$
 which is indeed the photon number difference between the two modes of the field. These states, sometimes have been called pair coherent states, have been used for
 the description of the production of pions \cite{Bhaumik2} and other problems in quantum field theory \cite{Klauder}. The explicit form of $|\alpha,q\rangle$,
 represents a well-known class of states within the general theory of coherent states, reads as
 \begin{eqnarray}\label{alfa,q}
|\alpha,q\rangle=N_{q}(|\alpha|^{2})^{-1/2}\sum_{n=0}^{\infty}\frac{\alpha^{n}}{\sqrt{n! [n+|q|]!}}\left|n+\frac{q+|q|}{2},n-\frac{q-|q|}{2}\right\rangle,
\end{eqnarray}
 where $\alpha\in\mathbb{C}$, the kets $|m,n\rangle$ are the two-mode number states and $N_{q}$ is an appropriate normalization factor may be determined.
 The states in (\ref{alfa,q}) include two distinct sets of charge coherent states corresponding to $ q\geq 0$ and
 $ q\leq 0$. In (\ref{alfa,q}) the following definition has been assumed
 \begin{eqnarray}\label{cnQg}
\quad[n+|q|]!\doteq(1+|q|) (2+|q|) ... (n+|q|).
\end{eqnarray}
  An experimental scheme for generation of the states in (\ref{alfa,q}) has been proposed by Agarwal \cite{Agarwal1,Agarwal2}.
  Recently, by using the nonlinear coherent states method \cite{Vogel, Man'ko} one of us with his co-author have introduced the nonlinear charge
  coherent states in a general structure which will be briefly discussed here \cite{Eftekhari}. Consider $f$-deformed ladder operators
 \begin{eqnarray}\label{ABf}
 \hat{A}=\hat{a}f(\hat{n}_{a}),\quad\quad\hat{A^{\dag}}=f^{\dag}(\hat{n}_{a})\hat{a^{\dag}},\nonumber\\
 \hat{B}=\hat{b}f(\hat{n}_{b}),\quad\quad\hat{B^{\dag}}=f^{\dag}(\hat{n}_{b})\hat{b^{\dag}},
\end{eqnarray}
 where $\hat{a}(\hat{b})$, $\hat{a^{\dag}}(\hat{b^{\dag}})$ and $\hat{n}_{a}=\hat{a^{\dag}}\hat{a}\;(\hat{n}_{b}=\hat{b^{\dag}}\hat{b})$,
 are respectively bosonic annihilation, creation and number operators of mode $a(b)$, and $f(n)$ is an operator-valued
 function of intensity of radiation field (from now on has been assumed to be real) characterizes the nonlinearity nature of physical systems.
 The pair $f$-deformed annihilation operator $\hat{A}\hat{B}$ commutes with the charge operator, i.e., $[\hat{Q},\hat{A}\hat{B}]=0$. Thus,
 the latter two operators should satisfy the following eigenvalue equations
 \begin{eqnarray}\label{aQ2}
\;\;\hat{Q}|\xi,q,f\rangle = q|\xi,q,f\rangle,
\end{eqnarray}
\begin{eqnarray}\label{aAB2}
\hat{A}\hat{B}|\xi,q,f\rangle = \xi|\xi,q,f\rangle,\quad\quad\quad\quad\quad\xi\in\mathbb{C},
\end{eqnarray}
where $\hat{Q}$ and $q$ keep their previous definitions expressed
in (\ref{Q,BAR}). These eigenstates have in
 general two distinct representations which can be put in a single expression as follows:
\begin{eqnarray}\label{vijetabeQAB}
 |\xi,q,f\rangle= N(|\xi|^{2})^{-1/2}\sum_{n=0}^{\infty}\frac{\xi^{n}}{\sqrt{n![n+q]!}[f(n)]![f(n+q)]!}\left|n+\frac{q+|q|}{2},n-\frac{q-|q|}{2}\right\rangle,\nonumber\\
\end{eqnarray}
with the normalization constant given by
\begin{eqnarray}\label{cnQAB}
N(|\xi|^{2})=\sum_{n=0}^{\infty}\frac{|\xi|^{2n}}{n![n+|q|]!([f(n)]![f(n+|q|)]!)^{2}}.
\end{eqnarray}
Note that, in obtaining (\ref{vijetabeQAB}) and (\ref{cnQAB}) we have used the conventional definitions
\begin{eqnarray}\label{fact1}
f(n)]!\doteq  f(n) f(n-1) f(n-2) \cdots f(1), \qquad [f(0)]!\doteq 1,
\end{eqnarray}
and
\begin{eqnarray}\label{fact2}
[f(n+|q|)]!&\doteq & f(n+|q|) f(n-1+|q|) f(n-2+|q|)\cdots f(1+|q|), \\ \nonumber \qquad[f(|q|)]!&\doteq& 1.
\end{eqnarray}
On the other hand, Fan et al \cite{Fan1} introduced $|q,\lambda\rangle$ \cite{Fan2}, i.e., the common
eigenstate of $\hat{Q}$ defined in (\ref{Q,BAR}) and (Hermitian)
$\hat{g}$ operator is  given by
\begin{eqnarray}\label{g}
 \hat{ g}=(\hat{ a}+\hat{ b}^\dag) (\hat{ a}^\dag+\hat{ b}),
 \end{eqnarray}
 with eigenvalues $q$ and $\lambda$, respectively, i.e.,
\begin{eqnarray}\label{aQ3}
\hat{ Q}|q,\lambda\rangle=q|q,\lambda\rangle,
\end{eqnarray}
\begin{eqnarray}\label{ag1}
\hat{ g}|q,\lambda\rangle=\lambda|q,\lambda\rangle,\qquad\lambda\geq \ 0,
\end{eqnarray}
by using $\langle\eta|$ representation. This is due to the fact that $[\hat{Q},\hat{g}]=0$. The explicit form of
$|q,\lambda\rangle$ in two-mode Fock-space, has been deduced by the authors
reads as
\begin{eqnarray}\label{q,land}
|q,\lambda\rangle=e^{-\frac{\lambda}{2}}\sum_{n=max(0,-q)}^{\infty}H_{n+q,n}(\sqrt{\lambda},\sqrt{\lambda})\frac{1}{\sqrt{(n+q)!n!}}|n+q,n\rangle,
\end{eqnarray}
where $H_{m,n}$ is the two-variable Hermite polynomial, has been defined as
\begin{eqnarray}\label{Hm,n}
H_{m,n}(z,z^{*})=\sum_{k=0}^{min(m,n)}\frac{(-1)^{k}m!n!}{k!(m-k)!(n-k)!}z^{m-k}z^{*n-k}.
\end{eqnarray}
 In the present paper, our main aims may be expressed as follows: i) reobtaining the explicit form of $|q,\lambda\rangle$ in
 two-mode Fock-space by using a rather different method other than the $\langle\eta|$ representation has been followed in \cite{Fan1},
 ii) generalizing $\hat{g}$ operator (combination of bosonic annihilation and creation operators in (\ref{g})) to $\hat{G}$ operator
 (combination of $f$-deformed ladder operators) and obtaining the common eigenstates of $\hat{Q}$ and $\hat{G}$ have been called
 by us as $"f$-deformed charge coherent states$"$\footnote[2]{We have selected this name to be distinguished from the previous
 nonlinear charge coherent states in \cite{Eftekhari}, whereas it is clear that both classes of states are in fact nonlinear or $f$-deformed.}
 and finally iii) investigating some of the nonclassical features and quantum statistical properties of the $f$-deformed charge coherent
 states associated with a few quantum systems with particular nonlinearity functions, in addition to the state $|q,\lambda\rangle$ which
 is indeed a special case of our $f$-deformed charge coherent states with $f(n)=1$. Obviously, our new type of $f$-deformed charge coherent states
 in the present paper is substantially different from the states in (\ref{vijetabeQAB}) have been introduced in \cite{Eftekhari}. \\
 The paper is organized as follows. In Sec. 2, we will reobtain the two-mode (linear) charge coherent states, which is the common eigenvector
 of $\hat{Q}$ and $\hat{g}$ operators. Then, in Sec. 3, we will generalize the creation and annihilation operators to the $f$-deformed operators
 and construct the $f$-deformed charge coherent states. Next, in Sec. 4, as some physical realizations of the formalism we consider a few particular
 nonlinearity functions and then we study some of the nonclassical features and quantum statistical properties of the two-mode (linear) charge
 coherent states and $f$-deformed charge coherent states associated with those physical systems. At last,  in Sec. 5, we will present a summery and conclusion.


 \section{Two-mode (linear) charge coherent state: common eigenstate of $\hat{Q}$ and $\hat{g}$ operators}
 Two-mode (linear) charge coherent state are common eigenstate of charge operator introduced in (\ref{Q,BAR}) and $\hat{g}$
 operator defined in (\ref{g}) respectively with eigenvalues $q$ and $\xi$ have been expressed in (\ref{aQ3}) and (\ref{ag1}).
 It is worthwhile noticing that we start our discussion by imposing a little modification in the notation of Ref. \cite{Fan1}
 by changing the eigenstate $|q,\lambda\rangle$ to $|\xi,q\rangle$. The form of the state in two-mode Fock-space is considered to be
\begin{eqnarray}\label{xi,q}
|\xi,q\rangle=\sum_{n=0}^{\infty}\sum_{m=0}^{\infty}c_{n,m}|n,m\rangle.
\end{eqnarray}
Substituting (\ref{xi,q}) in (\ref{aQ3}), one readily finds
\begin{eqnarray}\label{n,m,q}
n=m+q .
\end{eqnarray}
With the latter result in mind, the explicit form of the state may be rewritten as
\begin{eqnarray}\label{xi,q+}
|\xi,q\rangle ^{(+)}=\sum_{n=0}^{\infty}c_{n+q,n}^{(+)}|n+q,n\rangle ,\qquad q\geq\ 0,
\end{eqnarray}
\begin{eqnarray}\label{xi,q-}
|\xi,q\rangle^{(-)}=\sum_{n=0}^{\infty}c_{n,n-q}^{(-)}|n,n-q\rangle ,\qquad q\leq 0.
\end{eqnarray}
Now by substituting (\ref{xi,q+}), for instance, into (\ref{ag1}) we find the recursion relation
\begin{eqnarray}\label{rb}
c_{n+q,n}^{(+)} &=& \frac{1}{\sqrt{(n+q)n}}\{c_{n+q-1,n-1}^{(+)}[\xi-(n+q)-(n-1)] \nonumber \\ &-& c_{n+q-2,n-2}^{(+)}\sqrt{(n+q-1)(n-1)}\}.
\end{eqnarray}\nonumber
\begin{eqnarray}\label{rb2}
\qquad\qquad\qquad\qquad\qquad\qquad\qquad\qquad\qquad\qquad\qquad\qquad
\end{eqnarray}
By straightforward but lengthy procedure the expansion
coefficients are then obtained in terms of $c_{q,0}^{+}$ as
follows
\begin{eqnarray}\label{cnQg1}
c_{n+q,n}^{(+)}=c_{q,0}^{(+)}\sqrt{[n+q]! n! }\sum_{k=0}^{n}\frac{(-1)^{k} \xi ^{n-k}}{k! [n+q-k]! (n-k)!},
\end{eqnarray}
and thus the explicit form of the state for $q\geq \ 0$ may be expressed as
\begin{eqnarray}\label{xi,q+2}
|\xi,q\rangle^{(+)}=c_{q,0}^{(+)}\sum_{n=0}^{\infty}\sqrt{[n+q]! n! }\sum_{k=0}^{n}\frac{(-1)^{k} \xi^{n-k}}{k! [n+q-k]! (n-k)!}|n+q,n\rangle.
\end{eqnarray}
Similar procedure can be performed for $q\leq 0$, which led us
to the explicit form of (linear) charge coherent state for
$q\leq0$ as
\begin{eqnarray}\label{xi,q-2}
|\xi,q\rangle^{(-)}=c_{q,0}^{(-)}\sum_{n=0}^{\infty}\sqrt{n![n-q]!}\sum_{k=0}^{n}\frac{(-1)^{k} \xi^{n-k}}{k! [n-q-k]! (n-k)!}|n,n-q\rangle .
\end{eqnarray}
In (\ref{xi,q+2}) and (\ref{xi,q-2}) $\xi\in\mathbb{C}$, and the
normalization constants can be easily calculated from:
\begin{eqnarray}\label{cnQg2}
c_{q,0}^{(\pm)}=\left[\sum_{n=0}^{\infty}[n+|q|]! n! \left(\sum_{k=\circ}^{n}\frac{(-1)^{k} \xi^{n-k}}{k! [n+|q|-k]! (n-k)!}\right)^2\right]^{-1/2}.
\end{eqnarray}
By setting $q\geq 0$ and $q\leq 0$, in (\ref{cnQg2}) one obtains
the exact form of normalization factors of states in
(\ref{xi,q+2}) and (\ref{xi,q-2}), respectively. It is worth
mentioning that in deriving the relations
(\ref{cnQg1})-(\ref{cnQg2}) we have used the following definition
\begin{eqnarray}\label{fact3}
[n+|q|-k]!&\doteq& (1+|q|) (2+|q|) \cdots (n+|q|-k)\nonumber \\
     &=& \frac{(n+|q|-k)!}{|q|!},
\end{eqnarray}
which is a slightly different from the  conventional definition of
$[f(n)]!$ frequently has been used in the literature devotes with
$f$-deformed coherent states. Noticing that the state in
(\ref{q,land}) has been introduced in \cite{Fan1} is unnormalized,
it is easy to check that our final states in (\ref{xi,q+2}) and
(\ref{xi,q-2}) is in exact consistence with (\ref{q,land}), when
one sets $\xi=\xi^{*}=\sqrt{\lambda}$ in (\ref{xi,q+2}) and
(\ref{xi,q-2}) with $\lambda\in\mathbb{R}$. Also, note that the
domain of the states obtained by us is the entire space of complex
plane, while in Ref. \cite{Fan1} $\lambda$ is real and positive.
We would like to mention that our states, were obtained by a
simpler and at the same time more general manner, are normalized to
unity, while the state introduced in (\ref{q,land}) are not
normalized. In addition to this fact, to the best of our knowledge,
the statistical properties and nonclassical features of these
states have not been yet discussed in the literature. Henceforth,
we will pay attention to this subject in the continuation of the
paper, too, since any generalization scheme without this
discussion seems to be poor from physical points of view.


 \section{Introducing the $f$-deformed charge coherent states}\label{weight}
In this section, using the nonlinear coherent states method and
based on the procedure by which the charge coherent states are
derived in the previews section, we firstly generalize the bosonic
creation and annihilation operators to the $f$-deformed operators
defined in (\ref{ABf}) and therefore $\hat{g}$ which has been defined in
(\ref{g}) will be converted to $\hat{G}$ which is given by
 \begin{eqnarray}\label{G}
\hat{ G}=(\hat{A}+\hat{B}^\dag)(\hat{A}^\dag+\hat{B}).
\end{eqnarray}
Now, we introduce $f$-deformed charge coherent states as simultaneous eigenstates of
charge operator in (\ref{Q,BAR}) and $\hat{G}$ in (\ref{G}) with eigenvalues $q$ and $\xi$, respectively, i.e.,
 \begin{eqnarray}\label{aQ}
 \hat{Q}|\xi,q,f\rangle=q|\xi,q,f\rangle,
 \end{eqnarray}
 \begin{eqnarray}\label{aG}
 \hat{ G}|\xi,q,f\rangle=\xi|\xi,q,f\rangle,
 \end{eqnarray}
 since it can be easily checked that $[\hat{Q},\hat{G}]=0$
 $"$\footnote[2]{It is worth mentioning that we can also deform the bosonic charge $Q$ to
 $f$-deformed charge operator, however it then will not commute with
 neither of the operators $\hat a \hat b$,  $\hat A \hat B$, $\hat g$ or $\hat G$. So, no
 common eigenstate may be expected.}.
 The explicit form of the common eigenstate in two-mode Fock-space is described as
\begin{eqnarray}\label{Fock-space}
|\xi,q,f\rangle=\sum_{n=0}^{\infty}\sum_{m=0}^{\infty}c_{n,m}|n,m\rangle.
\end{eqnarray}
Substituting (\ref{Fock-space}) in (\ref{aQ}), one obtains the same expression as in (\ref{n,m,q}). With this result in mind, the explicit form of the state can be rewritten as
\begin{eqnarray}\label{vijetabeQG1}
|\xi,q,f\rangle^{(+)}=\sum_{n=0}^{\infty}c_{n+q,n}^{(+)}|n+q,n\rangle ,\qquad q\geq\ 0,
\end{eqnarray}
\begin{eqnarray}\label{vijetabeQG2}
|\xi,q,f\rangle^{(-)}=\sum_{n=0}^{\infty}c_{n,n-q}^{(-)}|n,n-q\rangle , \qquad q\leq\ 0 ,
\end{eqnarray}
where the superscript $+(-)$ indicates the positive (negative)
values of $q$ parameter. Now, substituting the state
(\ref{vijetabeQG1}), for instance, into (\ref{aG}) one finds a
complicated recursion relation for $q\geq0$ as follows:
\begin{eqnarray}\label{cn+q,n+2}
c_{n+q,n}^{(+)}&=&\frac{1}{\sqrt{(n+q)n}f(n+q)f(n)}\nonumber\\&\times&[c_{n+q-1,n-1}^{(+)}
(\xi-(n+q)f^{2}(n+q)-(n-1)f^{2}(n-1))\nonumber\\&-&c_{n+q-2,n-2}^{(+)}(\sqrt{(n+q-1)(n-1)})f(n+q-1)f(n-1))].
\end{eqnarray}
However, by defining
\begin{eqnarray}\label{Bn,Bn-1}
B_{n}^{(+)}=\frac{c_{n+q,n}^{(+)}}{c_{n+q-1,n-1}^{(+)}} \\
B_{n-1}^{(+)}=\frac{c_{n+q-1,n-1}^{(+)}}{c_{n+q-2,n-2}^{(+)}},
\end{eqnarray}
the relation (\ref{cn+q,n+2}) can be inverted to the form
\begin{eqnarray}\label{Bn+1}
B^{(+)}_{n}&=\frac{1}{\sqrt{(n+q)n}f(n+q)f(n)}\{[\xi-(n+q)f^{2}(n+q)-(n-1)f^{2}(n-1)]\nonumber\\
&-\frac{\sqrt{(n+q-1)(n-1)}f(n+q-1)f(n-1)}{B^{(+)}_{n-1}}\}.
\end{eqnarray}
From the above equation (\ref{Bn+1}), we may obtain an explicit
expression for $B_{n-1}^{(+)}$ in terms of $B_{n-2}^{(+)}$.
Setting the obtained $B_{n-1}^{(+)}$ in (\ref{Bn+1}) we arrive at
$B_{n}^{(+)}$ in terms of $B_{n-2}^{(+)}$. Continuing this
procedure we can obtain  $B_{n}^{(+)}$ in terms of
$B_{n-3}^{(+)}$, $\cdots$. Finally, we arrive at the complicated
expression for $B_{n}^{(+)}$ in terms of $B_{0}^{(+)}$ which is
assumed to be $1$:
\begin{eqnarray}\label{Bnq+}
B_{n}^{(+)}=\frac{1}{\sqrt{(n+q)n}f(n+q)f(n)}[[\xi-(n+q)f^{2}(n+q)-(n-1)f^{2}(n-1)] \nonumber\\
-\frac{(n+q-1)(n-1)f^{2}(n+q-1)f^{2}(n-1)}{[\xi-(n+q-1)f^{2}(n+q-1)-(n-2)f^{2}(n-2)]-\frac{(n+q-2)(n-2)f^{2}
(n+q-2)f^{2}(n-2)}{\vdots}}]\nonumber\\ \qquad\qquad\qquad\qquad\qquad\qquad\qquad
\qquad\qquad[\xi - (1+q)f^{2}(1+q)]\}.\nonumber
\end{eqnarray}
\begin{eqnarray}\label{khali2}
\qquad\qquad\qquad\qquad\qquad\qquad\qquad\qquad\qquad\qquad\qquad\qquad
\end{eqnarray}
Note that the following relations hold
\begin{eqnarray}\label{cbn}
 c_{n+q,n}^{(+)}=B_{n}^{(+)}c_{n+q-1,n-1}^{(+)}\nonumber\\
\quad\quad\;\;\ =B_{n}^{(+)}B_{n-1}^{(+)}c_{n+q-2,n-2}^{(+)}\nonumber\\
\quad\quad\;\;\ =B_{n}^{(+)}B_{n-1}^{(+)}B_{n-2}^{(+)}c_{n+q-3,n-3}^{(+)}\nonumber\\
\quad\quad\quad\qquad \vdots \nonumber\\
\quad\quad\;\;\ =B_{n}^{(+)}B_{n-1}^{(+)}B_{n-2}^{(+)}\;\; \ldots \;\;B_{0}^{(+)}c_{q,0}^{(+)},\;\;\;\;B_{0}^{(+)}\doteq 1,
\end{eqnarray}
or in a compact form we have:
\begin{eqnarray}\label{cn+q,n+3}
c_{n+q,n}^{(+)}=[B^{(+)}_{n}]! c_{q,0}^{(+)}.
\end{eqnarray}
Adding all of the above results, one readily deduces the explicit
form of $f$-deformed charge coherent states for $q\geq0$ as
follows:
\begin{eqnarray}\label{xi,q,f+}
 |\xi,q,f\rangle^{(+)}=c_{q,0}^{(+)}\sum_{n=0}^{\infty}[B^{(+)}_{n}]!|n+q,n\rangle .
\end{eqnarray}
The normalization constant in (\ref{xi,q,f+}) is given by
\begin{eqnarray}\label{cq,0+}
 c_{q,0}^{(+)}=\left(\sum_{n=0}^{\infty}\left|[B^{(+)}_{n}]!\right|^{2}\right)^{-1/2}.
\end{eqnarray}
Similar procedure may be followed for $q\leq0$, which leads one to the explicit form of $f$-deformed charge coherent states as
\begin{eqnarray}\label{xi,q,f-}
 |\xi,q,f\rangle^{(-)}=c_{q,0}^{(-)}\sum_{n=0}^{\infty}[B^{(-)}_{n}]!|n,n-q\rangle ,
\end{eqnarray}
where $B_{n}^{(-)}$ may be expressed as
\begin{eqnarray}\label{Bnq-}
B^{(-)}_{n}=\frac{1}{\sqrt{n(n-q)}f(n)f(n-q)}\{[\xi-(n)f^{2}(n)-(n-q-1)f^{2}(n-q-1)]\nonumber\\
-\frac{(n-1)(n-q-1)f^{2}(n-1)f^{2}(n-q-1)}{[(\xi-(n-1)f^{2}(n-1)-(n-q-2)f^{2}(n-q-2)]-\frac{(n-2)(n-q-2)f^{2}
( n-2)f^{2}(n-q-2)}{\vdots}}\nonumber\\ \qquad\qquad\qquad\qquad\qquad\qquad\qquad
\qquad\qquad[\xi-f^{2}(1)+(q)f^{2}(-q)]\}.\nonumber
\end{eqnarray}
\begin{eqnarray}\label{khali2}
\qquad\qquad\qquad\qquad\qquad\qquad\qquad\qquad\qquad\qquad\qquad\qquad
\end{eqnarray}
The normalization constant in (\ref{xi,q,f-}) can be simply obtained as:
\begin{eqnarray}\label{cq,0-}
 c_{q,0}^{(-)}=\left(\sum_{n=0}^{\infty}\left|[B^{(-)}_{n}]!\right|^{2}\right)^{-1/2}.
\end{eqnarray}
Notice that in all above formula we have replaced $ B^{(\pm)}_{n}(\xi, q, f)$ with  $ B^{(\pm)}_{n}$ for simplicity.
It can be easily checked that setting $f(n)=1$ in (\ref{xi,q,f+}) and (\ref{xi,q,f-}) recover (\ref{xi,q+2}) and (\ref{xi,q-2}), respectively.

 \section{Physical properties of the introduced states}
As is required, we briefly review some of the criteria which will
be used in the continuation of the paper for investigating the
nonclassicality of the introduced states, as Mandel parameter,
second-order correlation function, second-order correlation
function between the two modes, Cauchy-Schwartz inequality and finally the quasi-probability function $Q(\alpha)$.


 \subsection{Nonclassicality criteria}
  \begin{itemize}

  \item{} Using the definitions of position and momentum
      quadratures as $x=(\hat{a}+\hat{a}^\dag)/\sqrt{2}$ and
      $p=(\hat{a}-\hat{a}^\dag)/i\sqrt{2}$, it is well-known
      that the squeezing  respectively occurs in $x$ or $p$ if
      $(\triangle x)^{2}<0.5$ or $(\triangle p)^{2}<0.5$. For instance, for position quadrature one has
 \begin{eqnarray}\label{delta x}
  (\Delta x)^{2}&=&\langle x^{2}\rangle-\langle x \rangle^{2}
  \nonumber\\
   &=& \frac  1 2 [\langle \hat{a}^{2}\rangle
 +\langle {\hat{a}^{\dag ^2}}\rangle
 +\langle  \hat{a}^{\dag}\hat{a}\rangle
 +\langle \hat{a}\hat{a}^{\dag}\rangle
 -\langle \hat{a}\rangle^{2}
 -\langle\hat{a}^{\dag} \rangle^{2}
 -2\langle\hat{a} \rangle \langle\hat{a}^{\dag}\rangle].
 \end{eqnarray}
 By calculating the necessary mean values over the states
 $|\xi,q,f\rangle^{(+)}$ it is easily seen that
 $\langle\hat{a} \rangle =\langle\hat{a}^{\dag}\rangle=\langle
 \hat{a}^{2}\rangle =\langle {\hat{a}^{\dag ^2}}\rangle=0$; All of these arise  from the fact that $\langle m, n|m', n'\rangle =\delta _{m,m'}\delta _{n,n'}$.
 Therefore, $(\Delta x)^{2}=1/2[\langle
 \hat{a}^{\dag}\hat{a}\rangle +\langle
 \hat{a}\hat{a}^{\dag}\rangle]=\langle
 \hat{a}^{\dag}\hat{a}\rangle+1/2$; Similar calculations for
 momentum quadrature lead us to the same result for $(\Delta p)^{2}$. So, $(\Delta
 x)^{2}=(\Delta p)^{2}$ and this, clearly, irrespective of the value of $\langle
 \hat{a}^{\dag}\hat{a}\rangle$, immediately leads one
 to conclude that no quadrature squeezing may be expected.
 The same discussion can be followed for our states with
 $q<0$.

\item{}
  Commonly, photon-counting statistics of the quantum states is investigated by evaluating Mandel parameter has been defined as \cite{Mandel},
  \begin{equation}\label{Mandel}
   Q_{i}=\frac{\left\langle \hat{n_{i}}^{2}\right\rangle-\left\langle \hat{n_{i}}\right\rangle^{2}}{\left\langle \hat{n_{i}}\right\rangle}-1,
  \end{equation}
where $i$ stands for the two modes $a,b$ and so $ \hat{n_{a}}=\hat{a^{\dag}}\hat{a}, \hat{n_{b}}=\hat{b^{\dag}}\hat{b}$.
The states for which $Q_{a(b)}=0$, $Q_{a(b)}<0$ and $Q_{a(b)}>0,$ respectively corresponds to Poissonian (standard coherent states),
sub-Poissonian (nonclassical states) and super-Poissonian (classical states) statistics.

 \item{}
Even though there are quantum states in which supper-/sub-Poissonian statistical behavior is appeared simultaneously with bunching/antibunching effect, but this is not absolutely true \cite{exp}.
Therefore, to investigate bunching or antibunching effects, second-order correlation function, defined as \cite{Glauber}:
\begin{equation}\label{g2(0)}
g_{a}^{(2)}(0)= \frac{\langle{\hat{a}^{\dag^2}}\hat{a}^{2}\rangle}{\langle\hat{a}^{\dag}\hat{a}\rangle^{2}},
\end{equation}
is widely used. Depending on the particular nonlinearity
function $f(n)$, has been chosen for the construction of any
class of $f$-deformed charge coherent states,  $g^{2}(0)>1$
and  $g^{2}(0)<1$ respectively indicates bunching and
antibunching effects. The case $g^{2}(0)=1$ corresponds
particularly to the canonical coherent states.

\item{}
Second-order correlation function between the two modes of the radiation field has been defined as \cite{Song}
\begin{equation}\label{g12(0)}
g_{12}^{(2)}(0)=\frac{\langle\hat{a}^{\dag}\hat{a}\;\hat{b}^{\dag}\hat{b\rangle}}{\langle \hat{a}^{\dag}\hat{a}\rangle\langle\hat{b}^{\dag}\hat{b\rangle}}.
\end{equation}
This quantity shows that the two modes are correlated and the state is classical if $g_{12}^{(2)}(0)>1$; otherwise it is nonclassical.

\item{}
As another quantity, recall that the Cauchy-Schwartz inequality \cite{Song} has been defined as
\begin{equation}\label{I(0)}
I_{0}= \frac{[\langle{\hat{a}^{\dag ^2}}\hat{a}^2\rangle\langle{\hat{b}^{\dag ^2}}\hat{b}^2\rangle]^{1/2}}{\mid\langle \hat{a}^{\dag}\hat{a}\;\hat{b}^{\dag}\hat{b}\rangle\mid}-1<0.
\end{equation}
If this inequality is violated, then the state is nonclassical.

 It is to be mentioned that each of the above signs are
 sufficient, not necessary, for a state to be nonclassical, so
 we pay not more attention to other nonclassicality criteria.

 \item{}
    Different quasi-probability functions have been
    proposed in quantum optics \cite{Zubairy}. Among them
    we pay attention to the Husimi function
    $Q(\alpha)$ which is defined for the single mode quantum state
    $|\psi\rangle$ as $Q(\alpha)=|\langle \alpha| \psi \rangle|^2/\pi$, where $|\alpha\rangle$ is the canonical coherent states.
    This positive definite function  which is defined over the phase
    space can be constructed in the homodyne experiments
    \cite{homodyne}. By this function the quantum interference effects in phase space have been
    illustrated \cite{mundarain}.
    Seemingly, it is possible to generalize this definition to our two-mode introduced states as follows
\begin{eqnarray}\label{quasi-probability2}
Q(\alpha_{1},\alpha_{2} ,\xi)&=&\frac{1}{\pi}|\langle \alpha_{1},\alpha_{2} |\psi(\xi) \rangle|^{2},
\end{eqnarray}
where $\langle \alpha_{1},\alpha_{2} |$ is the bra of the ket $|\alpha_{1},\alpha_{2}\rangle$:
\begin{eqnarray}\label{alfa}
|\alpha_{1},\alpha_{2}\rangle&=&|\alpha_{1}\rangle\otimes|\alpha_{2}\rangle\nonumber\\
&=&e^{-\frac{|\alpha_{1}|^{2}}{2}}e^{-\frac{|\alpha_{2}|^{2}}{2}}\sum_{n_{1}=0}^{\infty}
\sum_{n_{2}=0}^{\infty}\frac{\alpha_{1}^{n_{1}}\alpha_{2}^{n_{2}}}{\sqrt{n_{1}!n_{2}!}}|n_{1},n_{2}\rangle.
\end{eqnarray}
Therefore, one can plot this function by fixing $\alpha_{1}$ or $\alpha_{2}$ in three-dimensional graphs.
\end {itemize}


\subsection{Some physical appearances of the introduced states }

In this section we want to investigate the nonclassical
properties of the reobtained two-mode (linear) charge coherent states and
the $f$-deformed charge coherent states. Clearly, choosing
different $f(n)$'s lead to a variety of states with various
physical properties.Therefore, one can not
investigate the physical properties of our introduced states
in (\ref{xi,q,f+}) and (\ref{xi,q,f-}) unless the special
physical system of interest is determined. In this work we confine ourselves to
three types of well-known nonlinearity functions.

i) Penson-Solomon nonlinearity function is defined by
$f_{PS}(n)=p^{1-n}$, where $0 \leq p \leq 1$. This function,
indeed, may be executed from the special class of coherent
states defined by Penson and Solomon \cite{PS} by using the
nonlinear coherent states method \cite{mktr}.

ii) A nonlinearity of the type $f(n)=\sqrt{n}$.
This function appears in a natural way in the Hamiltonian describing the interaction between a two-level atom and electromagnetic
field with intensity dependent coupling \cite{singh, obada}.

iii) The $\mathbf{q}$-deformed nonlinearity associated with the
 well-known $\mathbf{q}$-coherent states, defined as
 $f_{\mathbf{q}}(n)=\sqrt{\frac{\mathbf{q}^{n}-\mathbf{q}^{-n}}{n(\mathbf{q}-\mathbf{q}^{-1})}}$
 \cite{Man'ko}. Note that since charge is denoted usually by
 $q$, therefore we have used $\mathbf{q}$ notation in
 defining the $\mathbf{q}$-nonlinearity function.

Inserting each of these three nonlinearity functions in (\ref{xi,q,f+}) and (\ref{xi,q,f-}) one readily may obtain the explicit form of the associated $f$- deformed charge coherent states. \\
\begin{itemize}

 \item{} Evaluating the Mandel parameter for the
 $f$-deformed positive charge coherent states requires
 the following expectation values
 \begin{eqnarray}\label{ada}
  \langle \hat{n}_{a}\rangle^{(+)} &=&^{(+)}\langle\xi,q,f|\hat{n}_{a}|\xi,q,f\rangle^{(+)}\nonumber\\
  &=&|c_{q,0}^{(+)}|^{2}\sum_{n=0}^{\infty}\left|[B^{(+)}_{n}]!\right|^{2}(n+q),
 \end{eqnarray}
and
\begin{eqnarray}\label{ada2}
 \langle \hat {n}_{a}^{2}\rangle^{(+)} &=&^{(+)}\langle\xi,q,f|\hat {n}_{a}^{2}|\xi,q,f\rangle^{(+)}\nonumber\\
 &=&|c_{q,0}^{(+)}|^{2}\sum_{n=0}^{\infty}\left|[B^{(+)}_{n}]!\right|^{2}(n+q)^{2},
\end{eqnarray}
where $ \hat{n_{a}}=\hat{a^{\dag}}\hat{a}$.
In Fig. 1, the Mandel parameter for $f$-deformed charge
coherent states is shown for Penson-Solomon nonlinearity
function with $p=0.5$ and fixed charge  $q=1$, as well as
the $\mathbf{q}$-deformation function  with $\mathbf{q}=7$ and fixed
charge  $q=2$ .
It is seen that, for both cases,  this
parameter is always negative, and so the sub-Possonian
behavior is visible. Also, note that its value is nearly
$\simeq -1$, indicates the high strength nonclassicality
of the considered states. It ought to be mentioned that we have
also calculated the Mandel parameter for $f(n)=$ $\sqrt n$
and $f(n)=1$, but in both of the latter cases it gets
very high positive value relative to the previous ones,
such that displaying them in this figure would make the
graphs unclear. Moreover, we were sure about the lack of
this nonclassicality criteria for the other two functions.
\item{}
To calculate the second-order correlation function for mode $a$, defined in (\ref{g2(0)}), the following relation is needed:
\begin{eqnarray}\label{ad2a2}
 \langle {\hat{a}^{\dag^2}}\hat{a}^{2}\rangle ^{(+)}&=& ^{(+)}\langle\xi,q,f|{\hat{a}^{\dag ^2}}\;\hat{a}^{2}|\xi,q,f\rangle ^{(+)}\nonumber\\
&=&|c_{q,0}^{(+)}|^{2}\sum_{n=0}^{\infty}\left|[B^{(+)}_{n}]!\right|^{2}(n+q)(n+q-1),
\end{eqnarray}
and $\langle\hat{a}^{\dag}\hat{a}\rangle$ is also obtained in
(\ref{ada}).
From Fig. 2, it is visible that,  the inequality
$g^{(2)}(0)>1$ holds for two-mode (linear) charge coherent
states with $(f(n)=1)$, as well as the $f$-deformed charge
states with $f(n)=\sqrt n$ with respectively charge
parameters $q=1, 3$, which indicate the bunching effect
of the corresponding states.  Interestingly, the
$\mathbf{q}$-deformation plot shows that it becomes exactly 1,
like the canonical coherent states, where $q=1, \mathbf{q}=7$.
This criteria gets the values less than 1, only for the
Penson-Solomon nonlinearity function when we considered
charge a $q=-1$, i.e., it shows antibunching
(nonclassical) effect in a finite region of $\xi \in
\mathbb{R}$. Recalling that this nonclassicality sign
becomes $2$ for thermal light, we observe from figure 2
that for the cases $f(n)=\sqrt n$ and $f_{PS}(n)$ this
function takes values greater than $2$ in some regions of
$\xi$, i.e., in view of this criterion the corresponding states behave like a
supper-thermal light.
\item{} For the second-order correlation function between
    the two modes, the following mean values are
    necessary:
\begin{equation}\label{bdb}
\langle\hat{b}^{\dag}\hat{b}\rangle ^{(+)}=  ^{(+)}\langle\xi,q,f|\hat{b}^{\dag}\hat{b}|\xi,q,f\rangle^{(+)}
=|c_{q,0}^{(+)}|^{2}\sum_{n=0}^{\infty}n\left|[B^{(+)}_{n}]!\right|^{2},
\end{equation}
and
 \begin{eqnarray}\label{adabdb}
 \langle \hat{a}^{\dag}\hat{a}\;\hat{b}^{\dag}\hat{b}\rangle^{(+)} &=& ^{(+)}\langle\xi,q,f|\hat{a}^{\dag}\hat{a}\;\hat{b}^{\dag}\hat{b}|\xi,q,f\rangle^{(+)}\nonumber\\
 &=&|c_{q,0}^{(+)}|^{2}\sum_{n=0}^{\infty}\left|[B^{(+)}_{n}]!\right|^{2}n(n+q),
 \end{eqnarray}
 and $\langle\hat{a}^{\dag}\hat{a}\rangle$ is also obtained
 in (\ref{ada}).
 Fig. 3 shows $g_{12}^{(2)}(0)$ for two-mode linear and
 $f$-deformed charge coherent states. From the figure it
 is clear that again for the $\mathbf{q}$-deformation function
 with charge $q=2$ and $\mathbf{q}=7$  this criteria gets exact
 value 1, like the canonical coherent states; while for
 all other nonlinearity functions and the chosen charge
 parameters $q$  the value of $g_{12}^{(2)}(0)$ becomes
 greater than 1, indicating bunching (classical) effect of the corresponding field.

\item{} For Cauchy-Schwartz inequality, defined in
 (\ref{I(0)}), the following mean value is necessary:
 \begin{eqnarray}\label{bd2b2}
 \langle {\hat{b}^{\dag ^2}}\hat{b}^{2}\rangle^{(+)}&=&^{(+)}\langle\xi,q,f|{\hat{b}^{\dag ^2}}\;\hat{b}^{2}|\xi,q,f\rangle^{(+)}\nonumber\\
 &=&|c_{q,0}^{(+)}|^{2}\sum_{n=0}^{\infty}\left|[B^{(+)}_{n}]!\right|^{2}n(n-1),
\end{eqnarray}
and $\langle{\hat{a}^{\dag^2}}\hat{a}^{2}\rangle$ and
 $\langle\hat{a}^{\dag}\hat{a}\;\hat{b}^{\dag}\hat{b}\rangle$
are obtained in (\ref{ad2a2}) and (\ref{adabdb}),
respectively.
 Fig. 4 shows Cauchy-Schwartz inequality of  mode $a$ for
 two-mode linear and the three types of $f$-deformed
 charge coherent states. It is seen from the figure that
 in all cases this criterion gets negative value for the
 chosen charge parameters, indicating the nonclassicality
 feature of the associated states.

\item{}
    In addition, the probability of finding $n+q$
    photons in mode $a$ and $n$ photons in mode $b$ for
    our introduced positive charge states
    $|\xi,q,f\rangle^+$ reads as:
\begin{eqnarray}\label{p11}
  P^{(+)}(n+q,n)= |\langle n+q,n|\xi,q,f\rangle^{(+)}|^{2}&=&|c_{q,0}^{(+)}|^{2}\left|[B^{(+)}_{n}]!\right|^{2}.
\end{eqnarray}
 Fig. 5 shows photon number distribution for all of
 the considered nonlinearity functions including the
 linear one ($f(n)=1$). From the figure one may observe
 that, in all cases the oscillatory behaviour of the photon
 count is visible for the chosen $q$ and
 $\xi$ parameters. As is well-known this behaviour is one
 of the nonclassicality signs of the quantum states.
 Therefore, in view of this point, all of the considered
 charge states are nonclassical.

\item{}
 Quasi-distribution (Husimi) function for the  $f$-deformed
 charge coherent states with $q\geq 0$ read as
\begin{eqnarray}\label{p11}
Q(\alpha_{1},\alpha_{2} ,\xi)&=&\frac{1}{\pi}|\langle \alpha_{1},\alpha_{2}|\xi,q,f\rangle^{(+)}|^{2}\nonumber\\
&=&e^{-|\alpha_{1}|^{2}}e^{-|\alpha_{2}|^{2}}\sum_{n=0}^{\infty}\frac{|\alpha_{1}|^{2(n+q)}|\alpha_{2}|^{2(n)}}{n!(n+q)!}|c_{q,0}^{(+)}|^{2}\left|[B^{+}_{n}]!\right|^{2},
\end{eqnarray}
We have plotted the three-dimensional
graphs of quasi-probability distribution $Q(\alpha_{1},\alpha_{2})$ in terms of the amplitudes $x=\mathrm{Re}(\alpha_{1})$ and
 $y=\mathrm{Im}(\alpha_{1})$ for various nonlinearity
functions including $f(n)=1$ (linear) with fixed parameters $\xi$, $q$ and $\alpha_{2}$.
Generally, irrespective of the selected nonlinearity
functions, the plots corresponding to negative values of $q$ are similar, as
is the case for positive $q$'s. Specifically, from Fig. 6
it is obvious that,  in all cases, for the chosen parameters have been
used in our numerical calculations, when $q$
gets positive values there exists a hole (centered at the
origin of the complex plane) in the three-dimensional
graphs, while this hole will be disappeared when the $q$
parameter gets negative values.

\end{itemize}

We end this section with mentioning the following two points.
Firstly, although we brought the necessary mean values  for
our numerical calculations only for positive charge coherent
states $|\xi,q,f\rangle^{(+)}$, the calculation of similar
quantities for the $|\xi,q,f\rangle^{(-)}$ states, which have
been used in our numerical results may easily be done, too.
Secondly, our interpretations on the plotted graphs
specifically concern with the chosen fixed parameters have
been indicated in the related figure captions. So, obviously
the variation of the related parameters may lead one to enter states
with various physical behaviour and different nonclassicality features.

  \section{Summary and conclusion}
       In summary, we reobtained two-mode charge coherent
       states by a standard method rather than $\langle\eta|$
       representation. Then, we extended the latter approach
       to  nonlinear coherent states method and introduced the
       representation of two-mode $f$-deformed charge coherent
       states. The construction is valid for a large class of
       generalized nonlinear oscillators. However, in this
       paper we only used a few well-known nonlinearity
       functions associated with particular quantum systems as
       some physical appearances of our presented formalism.
       After introducing the explicit form of the associated
       states in two-mode Fock-space, we established that the
       $f$-deformed charge coherent state recovers two-mode
       charge coherent states when one sets $f(n)=1$. As a
       clear fact, it is observed that different $f(n)$'s lead
       to various classes of deformed charge coherent states,
       obviously with different physical properties, by tuning
       the parameters which exist in any case, i.e, $\xi$, $q$
       and sometimes like for the Penson-Solomon and
       $\mathbf{q}$-deformation an excess parameter, $p$ and $\mathbf{q}$,
       respectively. Anyway, due to the common motivation in
       the introduction of any generalized coherent states, i.e., investigating the nonclassicality features of the states, we
       pay attention to this matter by evaluating some of the
       nonclassicality properties of both two-mode linear and
       $f$-deformed charge coherent states.  As some criteria,
       we computed Mandel parameter, second-order correlation
       function, second-order correlation function between the
       two modes, Cauchy-Schwartz inequality, probability distribution function in addition to
       the quasi-probability Husimi function $Q(\alpha_{1},\alpha_{2})$ of the states, using
       three types of well-known nonlinearity functions,
       numerically. Summing up, all of the corresponding states have enough
       nonclassicality features to be classified in the
       nonclassical states, i.e., the states with no classical analogue.
       Although our work is mainly possesses
       mathematical-physics structure, we mention that the
       pair coherent states have found a few experimental
       schemes for their generations, so we hope that the
       two-mode charge coherent states in (\ref{xi,q+2}) and
       (\ref{xi,q-2}) which were originally introduced  in
       \cite{Fan1} and our $f$-deformed counterpart in
       (\ref{xi,q,f+}) and (\ref{xi,q,f-}) also find their
       appropriate experimental generation proposals in near
       future.

 \begin{flushleft}
  {\bf Acknowledgments}\\
 \end{flushleft}
 The  authors would like to thank  Dr M Hatami for his very
 useful helps in preparing computer programs for our numerical
 calculational results. Also, we thank from the referees for
 their time and helps which caused to improve and clarify
 the content of the paper.


 \newpage

 {\bf FIGURE CAPTIONS}

 \vspace {.5 cm}

{\bf FIG. 1} The plot of Mandel parameter for mode $a$ $(Q_{a})$, as a function of $\xi$ for $f$-deformed charge coherent states.
The dotted line is for $f_{PS}(n)$ with fixed parameter $p=0.5, q=1$ and the dashed line is for $f_{\mathbf{q}}$ with fixed parameter $q=2, \mathbf{q}=7$.

\vspace {.5 cm}

{\bf FIG. 2} The plot of second-order correlation function for
mode $a$ $(g_{a}^{2}(0))$, as a function of $\xi$, for
$f$-deformed charge coherent states. The continuous line is
for two mode (linear) charge coherent state $(f(n)=1)$ with
fixed parameter $q=1$, dotted line is for $f_{PS}(n)$ with
fixed parameter $q=-1, p=0.5$, dashed line is for $f_{\mathbf{q}}(n)$ with fixed
parameter $q=1, \mathbf{q}=7$ and the dot-dashed line is for $f(n)=\sqrt{n}$
with fixed parameter $q=3$.

 \vspace {.5 cm}

{\bf FIG. 3} The second-order correlation function between two modes $(g_{12}^{2}(0))$, as a function of $\xi$, for $f$-deformed charge coherent states. The continuous line is for two mode (linear) charge coherent state $(f(n)=1)$ with fixed parameter $q=-1$, dotted line is for $f_{PS}(n)$ with fixed parameter $q=-2, p=0.5$, the dashed line is for $f_{\mathbf{q}}(n)$ with fixed parameter $q=2, \mathbf{q}=7$ and the dot-dashed line is for $f(n)=\sqrt{n}$ with fixed parameter $q=1$.

\vspace {.5 cm}

{\bf FIG. 4} The plot of Cauchy-Schwartz inequality $(I_{0})$, as a function of $\xi$, for $f$-deformed charge coherent states. The continuous line is for two-mode (linear) charge coherent state $(f(n)=1)$ with fixed parameter $q=1$, dotted line is for $f_{PS}(n)$ with fixed parameter $q=1, p=0.5$, the dashed line is for $f_{\mathbf{q}}(n)$ with fixed parameter $q=3, \mathbf{q}=7$ and the dot-dashed line is for $f(n)=\sqrt{n}$ with fixed parameter $q=2$.

\vspace {.5 cm}

 {\bf FIG. 5} The plot of probability of finding $n+q$ photons in mode $a$ and $n$ photons in mode $b$ in $|\xi,q,f\rangle$, $(P(n+q,n))$, as a function of $n$, for $f$-deformed charge coherent states. The continuous line is for two-mode (linear) charge coherent state $(f(n)=1)$ with fixed parameter $\xi=5$ and $q=2$, dotted line is for $f_{PS}(n)$ with fixed parameter $\xi=10$ and $q=-1, p=0.5$, dashed line is for $f_{\mathbf{q}}(n)$  with fixed parameter $\xi=5$ and $q=-2, \mathbf{q}=7$ the dot-dashed line is for $f(n)=\sqrt{n}$ with fixed parameter $\xi=10$ and $q=1$.

\vspace {.5 cm}

{\bf FIG. 6} The plot of $Q(\alpha_1, \alpha_2)$ as a function of the amplitudes $x=\mathrm{Re}(\alpha_{1})$ and $y=\mathrm{Im}(\alpha_{1})$ for different classes of states.
 (a) two-mode (linear) charge coherent state $(f(n)=1)$, with fixed parameters $\xi=10$, $q=1$ and $\alpha_{2}=1+i$.
 (b) Two-mode (linear) charge coherent state $(f(n)=1)$, with fixed parameters $\xi=10$, $q=-1$ and $\alpha_{2}=1+i$.
(c)  $f$-deformed charge coherent state with $f_{PS}(n)$, fixed parameters are $\xi=10$, $q=2, p=0.5$ and $\alpha_{2}=1+i$.
(d)  $f$-deformed charge coherent state with $f_{PS}(n)$, fixed parameters are  $\xi=10$, $q=-2, p=0.5$ and $\alpha_{2}=1+i$.
(e) $f$-deformed charge coherent state with $f_{\mathbf{q}}(n)$, fixed parameters are $\xi=10$, $q=3, \mathbf{q}=7$ and $\alpha_{2}=1+i$.
(f)  $f$-deformed charge coherent state with $f_{\mathbf{q}}(n)$, fixed parameters are $\xi=10$, $q=-3, \mathbf{q}=7$ and $\alpha_{2}=1+i$.
(g)  $f$-deformed charge coherent state with $f(n)=\sqrt{n}$, fixed parameters are $\xi=10$, $q=4$ and $\alpha_{2}=1+i$.
 (h)  $f$-deformed charge coherent state with $f(n)=\sqrt{n}$, fixed parameters are $\xi=10$, $q=-4$ and $\alpha_{2}=1+i$.

\vspace {.5 cm}

 \end{document}